# LARGE-SCALE POWER SPECTRUM FROM PECULIAR VELOCITIES VIA LIKELIHOOD ANALYSIS

Saleem Zaroubi[1], Avishai Dekel[2], Yehuda Hoffman[2] & Tsafrir Kolatt[3]

[1] Astronomy Department and Center for Particle Astrophysics, University of California, Berkeley, CA 94720
[2] Racah Institute of Physics, The Hebrew University, Jerusalem 91904, Israel
[3] Harvard-Smithsonian Center for Astrophysics, 60 Garden St., Cambridge MA 02138

## ABSTRACT

The power spectrum (PS) of *mass* density fluctuations, independent of "biasing", is estimated from the Mark III Catalog of Peculiar Velocities of galaxies using Bayesian statistics. A parametric model is assumed for the PS, and the free parameters are determined by maximizing the posterior probability of the model given the data. The method has been tested using detailed mock catalogs. We use generalized CDM models with and without COBE normalization.

The robust result for all the models tested is $P(k)\Omega^{1.2} = (4.1 \pm 0.7) \times 10^3 \, (h^{-1}\mathrm{Mpc})^3$ at $k = 0.1 \, h \, \mathrm{Mpc}^{-1}$, with the peak constrained to the range $0.03 \leq k \leq 0.06 \, h \, \mathrm{Mpc}^{-1}$. It is consistent with a direct computation of the PS (Kolatt & Dekel 1996). When compared to galaxy-density surveys, the implied values for $\beta$ ($\equiv \Omega^{0.6}/b$) are of order unity to within 25%.

A $\Gamma$-shape model, free of COBE normalization, is constrained by the velocity data to $\Gamma = 0.5 \pm 0.15$ and $\sigma_8 \Omega^{0.6} = 0.85 \pm 0.1$. Within the family of COBE-normalized CDM models, the best tilted model ($\Omega = 1$, $n \approx 0.84 h_{50}^{-0.65}$) and the best open model ($n = 1$, $\Omega \approx 0.75 h_{50}^{-1.0}$) are more likely than the best $\Lambda$ model ($n = 1$, $\Lambda = 1 - \Omega$, $\Omega \approx 0.70 h_{50}^{-1.2}$). The most likely CDM model with $\Omega \leq 1$ is found to be of $\Omega = 1$, $h \approx 0.75$, and a *tilted* spectrum of $n = 0.8 \pm 0.02$ with tensor fluctuations. The tightest constraint for the tilted-$\Lambda$ models is of the sort $\Omega h_{50}^{1.2} n^\nu = 0.7 \pm 0.08$, with $\nu = 3.8$ and $1.85$ with and without tensor fluctuations respectively.

*Subject headings:* cosmology: theory — cosmology: observation — dark matter — galaxies: clustering — galaxies: distances and redshifts — large scale structure of universe

# 1. INTRODUCTION

In the standard picture of cosmology, the structure on large scales originated from small-amplitude initial density fluctuations that were amplified by gravitational instability. These initial fluctuations are assumed to be a Gaussian random field, solely characterized by it's power spectrum. On large scales, the fluctuations are linear even at late times, so that the power spectrum preserves it's original shape. This makes it a very useful statistics for large-scale structure.

The power spectra of galaxy density were derived for many different samples, in two angular dimensions or in three dimensions from redshift space. Unfortunately, these power spectra correspond to objects that are not necessarily unbiased tracers of the underlying mass distribution, and it is the mass distribution that is directly related to theory (*e.g.* Dekel & Rees 1987 for a review on "galaxy biasing"). Clear evidence for this bias is provided by the fact that galaxies of different types are observed to cluster differently (*e.g.* Dressler 1980). It would therefore be naive to assume that any of the galaxy power-spectra directly reflects the mass PS. Furthermore, in estimates of the galaxy PS from redshift surveys, uncertainties also arise when correcting for redshift distortions (Kaiser 1987, Zaroubi and Hoffman 1996). For these reasons, one wishes to measure the mass PS directly from dynamical data, bypassing the complex galaxy-biasing issues and the need to correct for redshift distortions. In principle, such dynamical information can be provided by peculiar velocities, by gravitational lensing effects, or by fluctuations in the cosmic microwave background (CMB). In particular, the accumulating catalogs of galaxy peculiar velocities enable a direct determination of the mass PS under the natural assumption that the galaxies are unbiased tracers of the large-scale, gravitationally-induced velocity field.

Here, the PS is computed from the Mark III Catalog of Peculiar Velocities (Willick *et al.* 1995 WI; 1996a WII; 1996b WIII), which consists of more than 3000 galaxies. It was compiled from several different data sets of spiral and elliptical/S0 galaxies with distances inferred by the forward Tully-Fisher and $D_n - \sigma$ methods, which were re-calibrated and self-consistently put together as a homogeneous catalog for velocity analysis. The catalog provides radial peculiar velocities and inferred distances, all properly corrected for inhomogeneous Malmquist bias, for $\sim 1200$ objects, ranging from isolated field galaxies to rich clusters. The associated errors are on the order of $17 - 21\%$ of the distance per galaxy. The sparse and inhomogeneous sampling is another source of error.

These data allow a reasonable recovery of the dynamical fields with $\sim 12\,h^{-1}{\rm Mpc}$ smoothing in a sphere of radius $\sim 60\,h^{-1}{\rm Mpc}$ about the Local Group, extending to $\sim 80\,h^{-1}{\rm Mpc}$ in certain regions. The POTENT method (Bertschinger & Dekel 1989; Dekel, Bertschinger & Faber 1990; Dekel 1994) attempts a recovery of the underlying density field with fixed Gaussian smoothing within this volume. In an associated paper, Kolatt and Dekel (1996, KD) have computed the mass PS from the smoothed density field recovered by POTENT from Mark III. The limitations of the data introduce severe systematic errors, that were modeled via Monte-Carlo mock catalogs and then used to correct the measured PS. Since the KD results naturally involve uncertainties, an independent estimate of the PS, using a very different method, is useful.

Our purpose is to estimate the mass PS directly from the peculiar velocities of the Mark III catalog, by means of a likelihood analysis. The non-local nature of the peculiar velocities, *i.e.* being influenced by the mass distribution in a whole neighborhood, allows one to probe scales somewhat larger than those probed by the density field. For example, the effect of a bulk velocity across the entire volume is not evident if only the density field is considered. For a similar reason, the velocity field is expected to obey linear theory better than the density field smoothed on a comparable scale, and so to be closer to a Gaussian field. Our approach here does not involve any explicit window function, weighting or smoothing, nor does it require artificial binning of the PS. In addition, it automatically underweights noisy, unreliable data.

The data analyzed here are especially suited for Bayesian analysis. The sparse and inhomogeneous sampling of a random Gaussian field with Gaussian errors yields a multivariate Gaussian data set. The corresponding *posterior* probability distribution function (PDF) is a multivariate Gaussian that is completely determined by the assumed PS and the assumed covariance matrix of errors. Under these conditions one can write the joint PDF of the model PS and the underlying velocity or density field, and then simultaneously estimate the PS model parameters and recover the "Wiener filter" solution of the fields (Zaroubi *et al.* 1995). In an associated paper (Zaroubi, Hoffman & Dekel 1996), we present the high-resolution fields recovered from this same data set using the PS derived here.

To apply our method, the simplifying assumptions that have to be made are that the peculiar velocities are drawn from a Gaussian field, that their correlation function can be derived from the density PS using linear theory, and that the errors are Gaussian. The need to assume a parametric functional form for the PS is also a limitation; one can try to achieve flexibility by using a large number of parameters and a variety of functional forms, but at the risk of sometimes making the likelihood analysis unstable (§5).

The method is described in §2, in which the relation between the velocity correlation functions and the PS, and the general likelihood-analysis algorithm for computing the PS, are specified. The method is tested using a mock catalog in §3. The resultant power spectra are presented in §4, as derived from the Mark III data alone, and for generalized CDM models imposing COBE normalization. The associated constraints on the cosmological parameters are analyzed. Our conclusions are summarized and discussed in §5.

## 2. METHOD

### 2.1. Velocity Correlations

The computation of the matter power spectrum from the peculiar velocity data by means of likelihood analysis requires a relation between the velocity correlation function and the power spectrum. Define the two-point velocity correlation ($3 \times 3$) tensor by the average over all pairs of points $\mathbf{r}_i$ and $\mathbf{r}_j$ that are separated by $\mathbf{r} = \mathbf{r}_j - \mathbf{r}_i$,

$$\Psi_{\mu\nu}(\mathbf{r}) \equiv \langle v_\mu(\mathbf{r}_i) v_\nu(\mathbf{r}_j) \rangle, \tag{1}$$

where $v_\mu(\mathbf{r}_i)$ is the $\mu$ component of the peculiar velocity at $\mathbf{r}_i$. In linear theory, it can be expressed in terms of two scalar functions of $r = |\mathbf{r}|$ (Górski 1988), parallel and perpendicular to the separation $\mathbf{r}$,

$$\Psi_{\mu\nu}(\mathbf{r}) = \Psi_\perp(r)\delta_{\mu\nu} + [\Psi_\parallel(r) - \Psi_\perp(r)]\hat{r}_\mu \hat{r}_\nu. \tag{2}$$

The spectral representation of these radial correlation functions is

$$\Psi_{\perp,\parallel}(r) = \frac{H_0^2 f^2(\Omega)}{2\pi^2} \int_0^\infty P(k)\, K_{\perp,\parallel}(kr)\, dk, \qquad (3)$$

where $K_\perp(x) = j_1(x)/x$ and $K_\parallel(x) = j_0 - 2j_1(x)/x$, with $j_l(x)$ the spherical Bessel function of order $l$. The cosmological $\Omega$ dependence enters as usual in linear theory via $f(\Omega) \approx \Omega^{0.6}$, and $H_0$ is the Hubble constant. A parametric functional form of $P(k)$ thus translates to a parametric form of $\Psi_{\mu\nu}$.

### 2.2. Likelihood Analysis

Let **m** be the vector of model parameters and **d** the vector of $N$ data points. Then Bayes' theorem states that the *posterior* probability density of a model given the data is

$$\mathrm{P}(\mathbf{m}|\mathbf{d}) = \frac{\mathrm{P}(\mathbf{m})\mathrm{P}(\mathbf{d}|\mathbf{m})}{\mathrm{P}(\mathbf{d})} \qquad (4)$$

The denominator is merely a normalization constant. The probability density of the model parameters, $\mathrm{P}(\mathbf{m})$, is unknown, and in the absence of any other information we assume it is uniform within a certain range. The conditional probability of the data given the model, $\mathrm{P}(\mathbf{d}|\mathbf{m})$, is the likelihood function, $\mathcal{L}(\mathbf{d}|\mathbf{m})$. The objective in this approach, which is to find the set of parameters that maximizes the probability of the model given the data, is thus equivalent to maximizing the likelihood of the data given the model (*cf.* Zaroubi *et al.* 1994; Jaffe & Kaiser 1994)

The Bayesian analysis measures the relative likelihood of different models. An absolute frequentist's measure of goodness of fit could be provided by the Chi-squared per degree of freedom, which we use as a check on the best parameters obtained by the likelihood analysis.

Assuming that the velocities are a Gaussian random field, the two-point velocity correlation tensor $\boldsymbol{\Psi}$ fully characterizes the statistics of the velocity field. Define the radial-velocity correlation ($N \times N$) matrix $U_{ij}$ by $U_{ij} = \hat{\mathbf{r}}_i^\dagger \boldsymbol{\Psi} \hat{\mathbf{r}}_j$, where $i$ and $j$ refer to the data points. Let the inferred radial peculiar velocity at $\mathbf{r}_i$ be $u_i$, with the corresponding error $\epsilon_i$ also assumed to be a Gaussian random variable. The observed correlation matrix is then $\tilde{U}_{ij} = U_{ij} + \epsilon_i^2 \delta_{ij}$, and the likelihood of the $N$ data points is

$$\mathcal{L} = [(2\pi)^N \det(\tilde{U}_{ij})]^{-1/2} \exp\left(-\frac{1}{2} \sum_{i,j}^N u_i \tilde{U}_{ij}^{-1} u_j\right). \qquad (5)$$

Given that the correlation matrix, $\tilde{U}_{ij}$, is symmetric and positive definite, we can use the Cholesky decomposition method (Press *et al.* 1992) for computing the likelihood function (Eq. 5). The significant contribution of the errors to the diagonal terms makes the matrix especially well-conditioned for decomposition. The calculation for a given choice of parameters and $N \sim 1200$ data points takes several minutes on a Dec-Alpha workstation (of SpecFP92 $\sim 150$).

The likelihood function of Eq. 5 is the posterior PDF of the parameters **m**. It is also a $\chi^2$ distribution (with $N$ degrees of freedom) with respect to the $N$ data points, but it is not necessarily a $\chi^2$ distribution with respect to the parameters, and it is therefore difficult to assign accurate confidence levels to the parameters. This requires elaborate integrations over the volume encompassed by the equal-likelihood surfaces in parameter space. In the present paper we limit ourselves to a rough estimate of confidence levels by crudely approximating the PDF as a $\chi^2$ distribution in parameter space.

Note that the quantity that can be derived from peculiar-velocity data via the linear approximation is $f^2(\Omega) P(k)$, where $P$ is the mass density PS (see Eq. 3).

## 3. TESTING THE METHOD

The Mark III catalog (WI; WII; WIII) provides inferred forward TF velocities to $\sim 1200$ objects, grouped from more than 3000 galaxies in order to reduce the distance uncertainty per object and thus reduce Malmquist bias. The velocities were further corrected for homogeneous and inhomogeneous Malmquist bias (*e.g.* Dekel 1994).

The grouping serves us here also as a means of smoothing over nonlinear velocities. The PS on large-scales should not be too affected by this grouping because the individual galaxies in clusters enter the likelihood with low weights anyway. This is because the randomness of the velocities within clusters makes the noise term $\epsilon_i^2 \delta_{ij}$ dominate over the signal $U_{ij}$ in the observed correlation matrix $\tilde{U}_{ij}$ in Eq. 5.

Careful testing of the method with realistic mock catalogs is essential in view of the large distance errors, the sparse and non-uniform sampling, the bias-correction procedures, and the possible non-linear and non-Gaussian effects.

The mock Mark III catalogs are described in Kolatt *et al.* (1996). They are based on simulations whose initial conditions were extracted from a reconstruction of the smoothed real-space density field from the IRAS 1.2Jy redshift survey, taken back into the linear regime. Small-scale perturbations were added by means of constrained random realizations. The system was then evolved forward in time using an N-body simulation assuming $\Omega = 1$, and stopped at two alternative times, when the *rms* density fluctuation in a top-hat sphere of radius $8\,h^{-1}$Mpc reached $\sigma_8 = 0.7$ and later when $sigma_8 = 1.12$. The "galaxies" in the simulation were identified via a linear biasing scheme (b=1.35), and they were divided into 'spirals' and 'ellipticals' according to Dressler's morphology-density relation. The galaxies were assigned TF quantities (internal velocities and absolute magnitudes) that were Gaussianly scattered about an assumed TF relation, and were then "observed" following the selection criteria of the actual data sets that compose the Mark III catalog. The mock catalogs were grouped and corrected for biases just like the real catalog.

The true PS of the mass in the simulation is well approximated by the functional form

$$P(k) = \frac{A_0\, k}{1 + (B\,k)^\alpha} , \qquad (6)$$

with $A_0 = 4.68$ and $12.28 \times 10^4 (\,h^{-1}\text{Mpc})^4$, $B = 8.3$ and $9.25\,h^{-1}$Mpc, and $\alpha = 3.2$ and 2.8, for the $\sigma_8 = 0.7$ and 1.12 cases respectively.

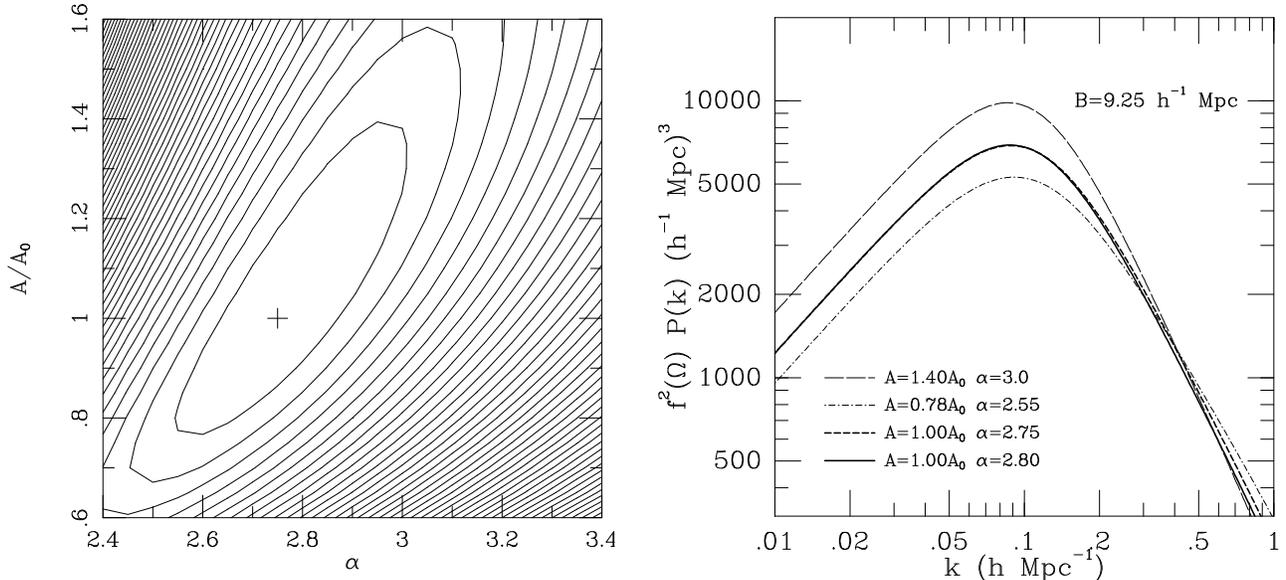

**Figure 1a:** Contour map of log likelihood in the $\alpha$–$A$ plane for a mock catalog based on the parametric model of Eq. 6 with $B = 9.25\,h^{-1}\mathrm{Mpc}$. Contour spacing is $\Delta[\ln(\mathcal{L})] = 1$. The best-fit point is marked.

**Figure 1b:** The true power spectrum of the simulation (heavy solid), compared with the best-fit solution (heavy dashed), and two power spectra whose parameters lie on the innermost closed contour of Fig. 1a. The values of $\sigma_8$ for this contour are $\approx 1 - 1.2$.

The likelihood analysis has been applied to the mock catalogs using the parametric functional form of Eq. (6) as a prior. To save effort we always kept one of the parameters fixed and allowed only the other two to vary. Figure 1a shows for example a contour map of $\ln(\mathcal{L})$ for one of the $\sigma_8 = 1.12$ mock catalogs, spanning the $\alpha$-$A$ plane with $B = 9.25\,h^{-1}\mathrm{Mpc}$. The contours are separated by $\Delta \ln \mathcal{L} = 1$. Maximum likelihood is obtained at $A = A_0$ and $\alpha = 2.75$ (compared to 2.80). Assuming a Chi-square distribution with two degrees of freedom, the $1\sigma$ contour of the likelihood around the best-fit parameters is at $\ln(\mathcal{L}) \approx 2.1$. We conclude that the recovery of the PS is excellent.

Figure 1b shows the recovered PS in comparison with the true PS of the simulation. They almost coincide over the whole range of scales, showing slight deviations only on very small scales. To illustrate the level of uncertainty, we plot for comparison several other power spectra that were obtained with parameter pairs that lie on the innermost contour about the maximum in Fig. 1a. It shows that the amplitude near the peak can be off by about 25%, and that the recovery becomes more robust at moderately smaller scales. The success of the recovery is similar when the other pairs of parameters are allowed to vary.

### 4. RESULTS

#### 4.1. The $\Gamma$ Model

To recover the PS from the velocity data independent of the COBE normalization, we use as a parametric prior the so-called $\Gamma$ model (*e.g.* Efstathiou, Bond and White 1992),

$$P(k) = A\,k\,T^2(k), \quad T(k) = \left(1 + [ak/\Gamma + (bk/\Gamma)^{3/2} + (ck/\Gamma)^2]^\nu\right)^{-1/\nu}, \qquad (7)$$

with $a = 6.4\,h^{-1}\mathrm{Mpc}$, $b = 3.0\,h^{-1}\mathrm{Mpc}$, $c = 1.7\,h^{-1}\mathrm{Mpc}$ and $\nu = 1.13$. The free parameters to be determined by the likelihood analysis are the normalization factor $A$ and the $\Gamma$ parameter. In the context of the CDM cosmological model, $\Gamma$ has a specific cosmological interpretation, $\Gamma = \Omega h$. Here, however, independently of CDM, Eq. 7 serves as a generic function with logarithmic slopes $n = 1$ and $-3$ on large and small scales respectively, and with a turnover at some intermediate wavenumber that is determined by the single shape parameter $\Gamma$.

Figure 2a shows the contour map of $\ln \mathcal{L}$ in the $\Gamma$-$A$ plane. The maximum likelihood values are $\Gamma = 0.5$ and $A = 9.4 \times 10^4 (h^{-1}\mathrm{Mpc})^4$. The corresponding value of $\sigma_8 \Omega^{0.6}$ is $0.85 \pm 0.1$. Figure 2b shows the best-fit PS (solid). To illustrate the uncertainty in the PS we also show the power spectra of five other parameter pairs that lie on the innermost likelihood contour about the best fit.

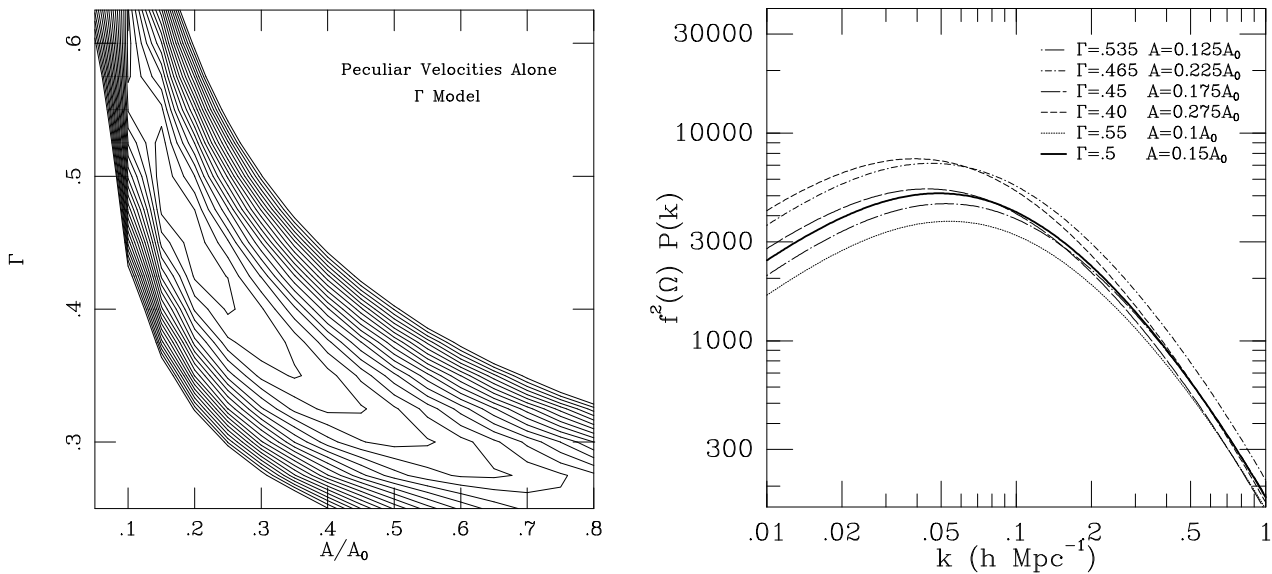

**Figure 2a:** Contour map of log likelihood for the $\Gamma$ model. Contour spacing is $\Delta[\ln(\mathcal{L})] = 1$. $A$ in units of $A_0 = 6.28 \times 10^5 (h^{-1}\mathrm{Mpc})^4$.

**Figure 2b:** The most likely $\Gamma$-model power spectrum (solid), and five other models whose parameters lie on the innermost contour of Fig 2a.

### 4.2. COBE-Normalized CDM Models

We now restrict our attention to the generalized family of CDM cosmological models, allowing variations in the cosmological parameters $\Omega$, $\Lambda$ and $h$, as well as the large-scale PS slope $n$ and the contribution of tensor fluctuations. Furthermore, we now impose the normalization implied by the two-year COBE DMR data as an additional external constraint. The general form of the PS in these models is

$$P(k) = A_{COBE}(n, \Omega, \Lambda)\, T^2(\Omega, \Omega_B, h; k)\, k^n, \qquad (8)$$

where we adopt the CDM transfer function proposed by Sugiyama (1995),

$$T(k) = \frac{\ln(1 + 2.3q)}{2.34q} \left[1 + 3.89q + (16.1q)^2 + (5.46q)^3 + (6.71q)^4\right]^{-1/4}, \qquad (9a)$$

$$q = k \left[ \Omega h \, \exp(-\Omega_b - h_{50}^{1/2} \Omega_b / \Omega) \, (h \, \text{Mpc}^{-1}) \right]^{-1} . \qquad (9b)$$

The parameters are varied, two at a time, such that they span the range of currently popular CDM models, including Tilted-$\Lambda$ CDM ($\Omega + \Lambda = 1$, $\Omega \leq 1$, $n \leq 1$) and Tilted-Open CDM ($\Lambda = 0$, $\Omega \leq 1$, $n \leq 1$). We allow the possibility of nonzero tensor fluctuations of $T/S = 7(1 - n)$, where the ratio is of the $C_2$ quadrupole moments of temperature angular fluctuations of the tensor and scalar modes (*e.g.* Turner 1993; Crittenden *et al.* 1993). In all cases, the baryonic density is assumed to be $\Omega_b = 0.0125 h^{-2}$, which is the value currently favored by primordial nucleosynthesis analysis (*e.g.* Walker *et al.* 1991).

The COBE normalization for each model has been calculated by various authors (Górski *et al.* 1995; Sugiyama 1995; White & Bunn 1995), using different Boltzmann codes, different statistical analyses, and sometimes even different temperature maps. We have arbitrarily adopted Sugiyama's normalization as a backbone, and for models not studied by him we use the other results after matching them to Sugiyama's using the models that they have investigated in common.

In particular, the COBE normalization is modeled by $A_{COBE} = A_1(\Omega) A_2(n)$. For Tilted-$\Lambda$ CDM models we use the fits:

$$A_1(\Omega) = \text{dex}(7.93 - 8.33\Omega + 21.31\Omega^2 - 29.67\Omega^3 + 10.65\Omega^4 +$$
$$15.42\Omega^5 - 6.04\Omega^6 + 13.97\Omega^7 + 8.61\Omega^8), \qquad (10a)$$

$$A_2(n) = \begin{cases} \text{dex}(-2.78 + 2.78n) & (T = 0); \\ \text{dex}(-4.54 + 4.54n) & (T \neq 0). \end{cases} \qquad (10b)$$

Thes fits are for $h = 0.5$, but the $h$ dependence in the range of interest is weak, and we ignore it here.

For Tilted-Open CDM model with $T = 0$ the fit is:

$$A_1(\Omega) = \text{dex}(5.75 - 1.68\Omega + 4.53\Omega^2 - 7.57\Omega^3 + 7.53\Omega^4 + 3.15\Omega^5 - 0.23\Omega^6), \qquad (11a)$$

$$A_2(n) = \text{dex}(-2.71 + 2.71n). \qquad (11b)$$

*4.2.1 Low Omega Models*

Figure 3a shows the likelihood contour map in the $\Omega - h$ plane, for the $\Lambda$CDM family of models with $n = 1$ (normalization by Sugiyama). The most probable parameters in this case (in the range $\Omega \leq 1$) are $\Omega = 1$ and $h = 0.4$. Figure 3b shows the similar likelihood map for OCDM with $n = 1$. The most probable values here are $\Omega = 0.46$ and $h = 0.9$. It is clear from the elongated contour maps that $\Omega$ and $h$ are not constrained very effectively independently of each other. It is a degenerate combination of the two parameters, approximately $\Omega h^x$ with $x \sim 1$ (*i.e.* a combination close to the $\Gamma$ parameter) that is being determined tightly by the elongated ridge of high likelihood.

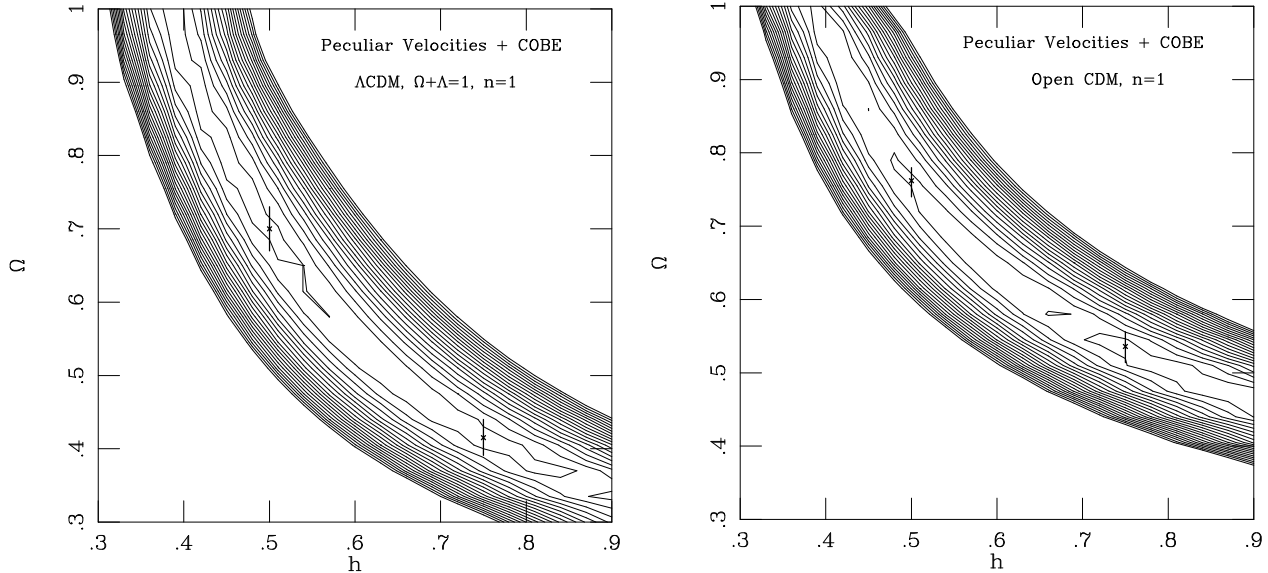

**Figure 3:** Contour map of log likelihood in the $\Omega - h$ plane for the $\Lambda$CDM (a) and OCDM (b) models with $n = 1$. $\Delta[\ln(\mathcal{L})] = 1$. The most likely values of $\Omega$ for two given values of $h$, and the corresponding $1\sigma$ error bars, are shown.

We can thus quote stringent constraints on the conditional best value of $\Omega$ given $h$: $\Omega \approx (0.7 \pm 0.06) h_{50}^{-1.2}$ for $\Lambda$CDM, and $\Omega \approx (0.75 \pm 0.06) h_{50}^{-1.0}$ for OCDM ($h_{50} \equiv H_0/50 \,\mathrm{km\,s^{-1}Mpc^{-1}}$).

Table 1 lists the most likely model and it's ln-likelihood (with the zero set arbitrarily) for each family of models, within the range $\Omega \leq 1$ and $n \leq 1$ and for two values of $h$. The errors are the $1\sigma$ conditional uncertainty. We can see from Fig. 2 and the table that, when $n = 1$ is enforced, the best open model is always more likely than the best $\Lambda$ model, for any given $h$.

*4.2.2 Tilted Models*

For the tilted $\Lambda$CDM family of models, with or without tensor fluctuations, we use the COBE normalization by White and Bunn (1995) (computed only for $h = 0.5$ and $0.75$). The results are presented in Figure 4. The best fits and their likelihood are listed in Table 1. The analysis prefers the highest $\Omega$ in the range, i.e. , $\Omega = 1$, with a nonzero tilt. It also prefers the higher value $h = 0.75$. Nonzero tensor fluctuations increase the likelihood. The best fit is obtained at $\Omega = 1$, $h = 0.75$, $n = 0.8$ and $T/S = 7(1 - n)$. The tight constraint is $\Omega h_{50}^{1.2} n^\nu \approx 0.7 \pm 0.08$, with $\nu = 1.85$ for $T/S = 0$, and $\nu = 3.8$ for $T/S = 7(1 - n)$.

For fixed $\Omega$, this relation can be understood qualitatively as follows: the normalization by COBE fixes the amplitude at small wavenumbers, $k \sim 0.001$, and the velocity data constrain the amplitude at $k \sim 0.1$. The wavenumber corresponding to the peak of the PS is proportional to $\Omega h$. Therefore, if a good fit is obtained with certain values of $h$ and $n$, a similarly good fit can be obtained with higher $h$ and lower $n$, or vice versa. The presence of tensor fluctuations lowers the amplitude imposed by COBE at small wavenumbers, and thus weakens the requirement for a tilt in $n$.

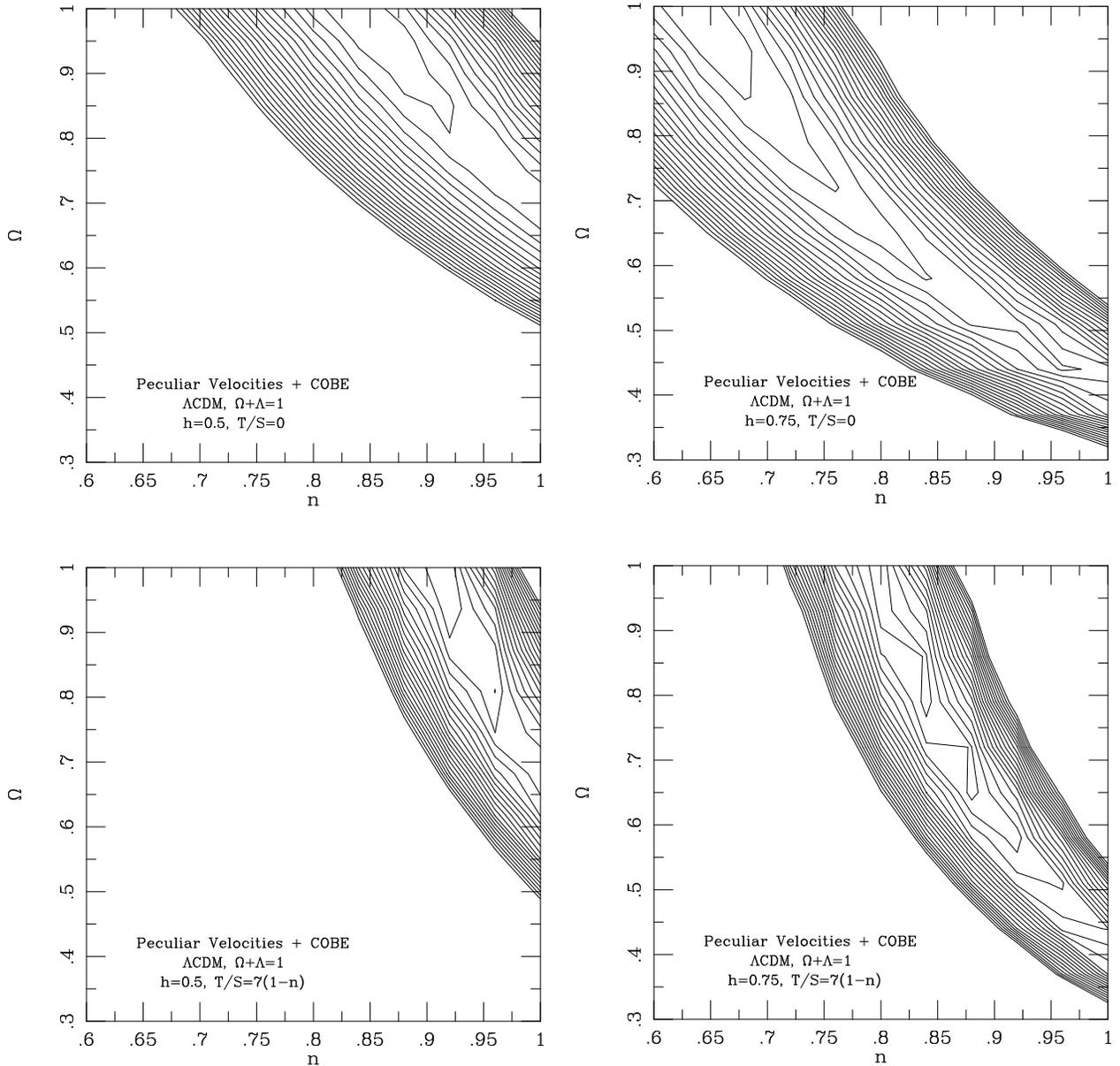

**Figure 4:** Contours of log likelihood in the $\Omega - n$ plane, calculated with $h = 0.5$ and $0.75$ for T$\Lambda$CDM models with and without tensor component. $\Delta[\ln(\mathcal{L})] = 1$.

The results for the tilted-open family of models (normalized by Sugiyama 1995), for the case $T/S = 0$ only, are presented in Figure 5. The tendency towards large $\Omega$ and $h$ is similar to the case of tilted-$\Lambda$ models, and the maximum likelihood is similar too, but the optimal tilt is more pronounced, $n = 0.64$. The tight constraint in this case is not very different either, $\Omega h_{50} n^{1.77} \approx (0.75 \pm 0.08)$.

The best-fit power spectra for the families of models discussed above are drawn in Figure 6. All the models agree to within $\sim 20\%$ for $k > 0.1 \, h \, \text{Mpc}^{-1}$, and they differ by up to $30 - 50\%$ on larger scales. Figure 7 shows that the most probable model from the COBE-normalized CDM models agrees with the $\Gamma$-model (that is COBE independent) on all scales to within $\sim 10\%$. This is a demonstration of the fact that the velocity data contain enough information to constrain the scale of the peak in the power spectrum even without the constraint from COBE. Fig. 7 also shows the typical scatter in the PS about the best CDM model, using parameter pairs that lie on the innermost likelihood contour (Fig. 4, lower-right panel). This scatter is somewhat smaller than in the $\Gamma$ model (Figure 2b) because of the additional constraint from COBE.

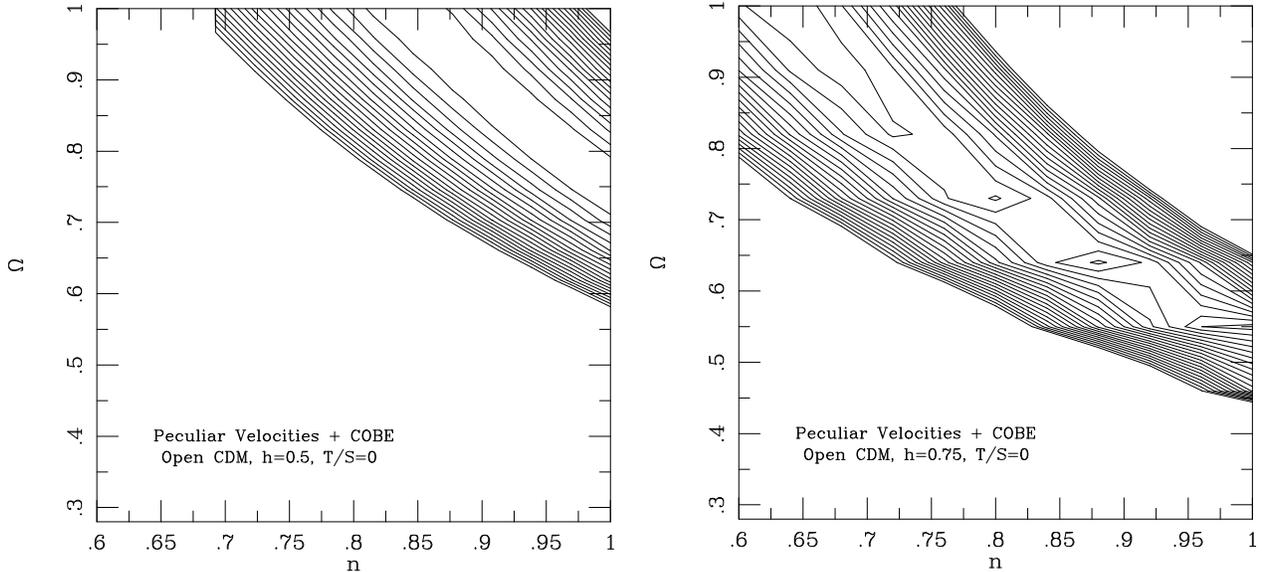

**Figure 5:** The same as Figure 4 for tilted Open CDM models without tensor component, $\Delta[\ln(\mathcal{L})] = 1$.

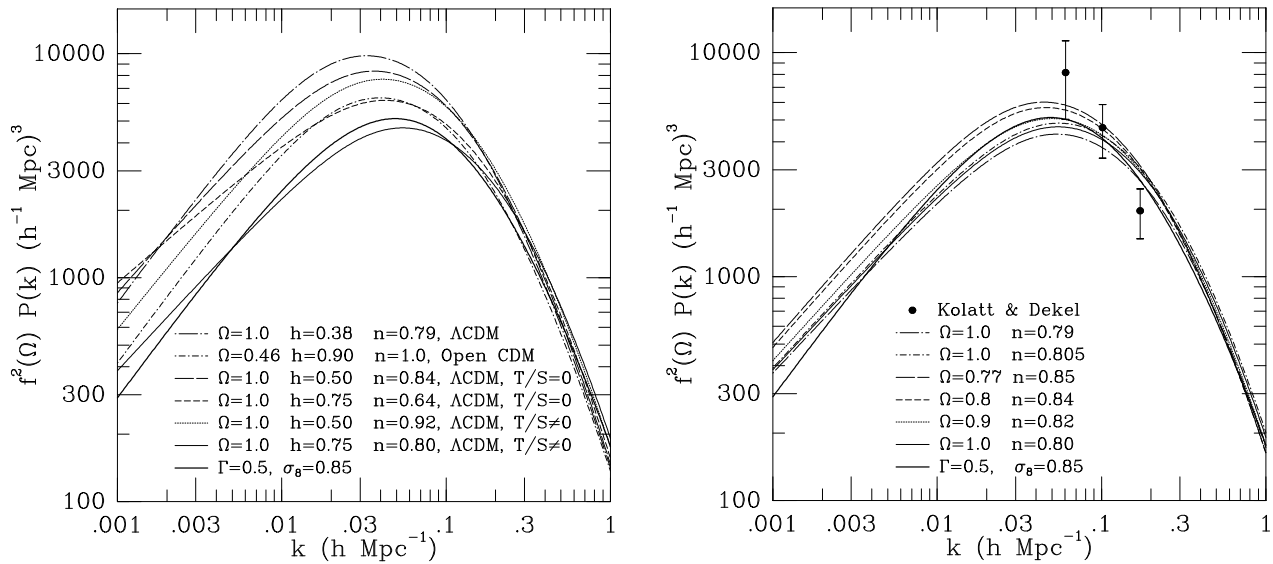

**Figure 6:** The best-fit power spectra for the various CDM models.

**Figure 7:** The PS of the most probable COBE-normalized CDM model (solid), and the scatter about it following parameter pairs that lie on the innermost likelihood contour (Fig. 4, lower-right panel). The COBE-free $\Gamma$ model is also marked (heavy solid). The PS computed by KD from POTENT density of the same velocity data (independent of COBE or models), and their measurement errors, are shown in three bins.

For comparison, Figure 7 also displays the PS as computed by KD from the POTENT smoothed density field recovered from the same Mark III data (independent of COBE or models). The results (for all the models tested here) agree within $1\sigma$ of the measurement errors, and they agree particularly well near $k = 0.1\,h\,\mathrm{Mpc}^{-1}$, where the velocity data imposes the strongest constraints. The KD spectrum seems to be somewhat steeper. In fact it is steeper than any of the CDM spectra discussed here (see Fig. 6), and is roughly as steep as the PS predicted by the CHDM model, a 3:7 mixture of hot and cold dark matter (KD). The result of the current paper is probably more reliable than KD on large scales, because the likelihood method uses all the velocity data including the large-scale flows, while the POTENT density field is insensitive to the bulk velocity. On the other hand, our results here may be less reliable on small scales because, other than grouping, we do not make any correction for nonlinear effects.

Table 1: CDM Models $\Omega \leq 1$, $n \leq 1$

| CDM Model | $\Omega$ | $\Lambda$ | $H$ | $n$ | $T/S$ | $\ln \mathcal{L}$ |
|---|---|---|---|---|---|---|
| Standard | 1 | 0 | 50 | 1 | 0 | $\ll 0$ |
|  | 1 | 0 | 40±2 | 1 | 0 | **2.15** |
| $\Lambda$ | 0.70±0.03 | $1-\Omega$ | 50 | 1 | 0 | **1.16** |
|  | 0.415±0.025 | $1-\Omega$ | 75 | 1 | 0 | **0.00** |
| Open | 0.74±0.02 | 0 | 50 | 1 | 0 | **2.71** |
|  | 0.54±0.02 | 0 | 75 | 1 | 0 | **3.66** |
| Tilted | 1 | 0 | 50 | 0.84±0.04 | 0 | **2.68** |
| $\Lambda$ or Open | 1 | 0 | 75 | 0.64±0.04 | 0 | **4.41** |
| Tilted+tensor | 1 | 0 | 50 | 0.91±0.03 | $7(1-n)$ | **3.71** |
|  | 1 | 0 | 75 | 0.80±0.02 | $7(1-n)$ | **5.31** |

## 5. CONCLUSION AND DISCUSSION

We have presented a Bayesian method for deriving the power spectrum of mass density fluctuations from the Mark III Catalog of Peculiar Velocities. The result is free of galaxy "biasing." The method extracts the maximum amount of useful information from the data. It is exact to first order, under the assumption of Gaussian fluctuations and Gaussian errors. Tests using realistic mock catalogs show that this approximation is adequate because of the large-scale coherence of the velocity field and because the large errors that dominate on small, nonlinear scales mean that nonlinear effects contribute only weakly to the result.

Our robust result for the whole family of models examined here is that the PS amplitude at $k = 0.1\,h\,\mathrm{Mpc}^{-1}$ is $P(k)\Omega^{1.2} = (4.1 \pm 0.7) \times 10^3\,(h^{-1}\mathrm{Mpc})^3$, with the PS peak in the range $0.03 \leq k \leq 0.06\,h\,\mathrm{Mpc}^{-1}$, which, for the best-fit model, translates to $\sigma_8 \Omega^{0.6} = 0.85 \pm 0.1$. The errors quoted are crude: they are the typical $1\sigma$ uncertainty for each of the best-fits within each family of models, and also the typical scatter among these best-fit models. Similar results are obtained when using the velocity data alone, as when the additional constraints from COBE are included. Moreover, the results are insensitive to the actual choice of model within the general family of CDM models. Note, however, that our quoted errors do not include the cosmic scatter that arises from the fact that we do not necessarily sample a fair sample of the universe. A crude estimate of the cosmic scatter in the PS for the standard CDM model can be found in KD (Fig. 8). It is comparable to, and even somewhat larger than, the measurement errors.

This normalization of the PS is in pleasant agreement with the independent computation of the PS by KD, which yielded $P(k)\Omega^{1.2} = (4.6 \pm 1.4) \times 10^3 \, (h^{-1}\mathrm{Mpc})^3$ at $k = 0.1 \, h\,\mathrm{Mpc}^{-1}$, and $\sigma_8 \Omega^{0.6} \approx 0.7 - 0.8$. Our new results differ at about the $2\sigma$ level from the earlier estimate by Seljak & Bertschinger (1994) based on the earlier Mark II sample, of $\sigma_8 \Omega^{0.6} = 1.3 \pm 0.3$. The main improvements since then are that the current analysis includes five times denser sampling in a more extended volume, the systematic errors such as Malmquist bias are handled better, and a wider span of models is used in the likelihood analysis. It may be interesting to note that the current measurement is somewhat higher than the completely independent estimate of a similar quantity based on cluster abundances, $\sigma_8 \Omega^{0.56} \simeq 0.57 \pm 0.05$ (White, Efstathiou, & Frenk 1993), but it is only about $2\sigma$ away.

The comparisons of the mass $\sigma_8$ to the values observed for optical galaxies ($\simeq 0.95$) and for IRAS 1.2Jy galaxies ($\simeq 0.6 - 0.7$) indicate $\beta$ values of order unity to within 25% for most galaxy types on these scales (see KD, Fig. 6 and Table 1, for more details).

A $\Gamma$-shape model, free of COBE normalization, is constrained by the velocity data to $\Gamma = 0.5 \pm 0.15$. The errors in $\sigma_8$ and $\Gamma$ are not independent; higher values of $\Gamma$ correspond to lower normalization, *i.e.* lower values of $\sigma_8 \Omega^{0.6}$. This value of $\Gamma$ is somewhat higher than the canonical values of $\sim 0.2 - 0.3$ typically obtained from galaxy density surveys (*e.g.* Efstathiou *et al.* 1992, Peacock & Dodds 1994).

Within the families of COBE-normalized CDM models, which we have restricted to the range $\Omega \leq 1$, the best-fit tilted models with $\Omega = 1$ [$n = (0.84 \pm 0.04)h_{50}^{-0.65}$], and the best-fit open models with $n = 1$ [$\Omega = (0.75 \pm 0.03)h_{50}^{-1.0}$], are found to be more likely than the best-fit $\Lambda$ models with $\Lambda = 1 - \Omega$ and $n = 1$ [$\Omega = (0.70 \pm 0.03)h_{50}^{-1.2}$].

Our analysis shows that the most likely among all the CDM models has $\Omega = 1$, $h \approx 0.75$, and a *tilted* spectrum of $n = 0.8 \pm 0.02$ with tensor fluctuations of $T/S = 7(1 - n)$. The most stringent constraints obtained using this likelihood analysis are of the sort $\Omega h_{50}^{\mu} n^{\nu} = 0.7 \pm 0.08$ with $\mu = 1.2$ for the $\Lambda$ models and with $\nu = 3.8$ and $1.85$ w/wo tensor fluctuations respectively. For the open models without tensor fluctuations it is $\mu = 1.0$ and $\nu = 1.77$.

Our results are consistent with the conclusion of White *et al.* (1995), who argue for tilted CDM based on several data sets including power spectra of galaxy density (Peacock & Dodds 1994), cluster correlations, pair-wise velocities and COBE's results. Their best fit is $\Omega = 1$, $h \approx 0.45$, $n = 0.9$, with tensor fluctuations. This model is about $1\sigma$ away from our best fit (see Fig. 6), but the PS is quite similar; our lower value of $n$ compensates for our higher value of $h$.

A note of caution about the method. If not enough constraints are imposed, the inversion of $\tilde{U}_{i,j}$ (Eq. 5) may be dominated by the noise rather than the signal. For example, when we tried to parameterize the PS with a function that did not enforce any upper bound on the power on large scales (as is properly enforced in the $\Gamma$ model, or when COBE constraints are used), then the likelihood analysis preferred unphysically high power on large scales. This is, at least in part, a result of noise dominance. An algorithm to detect and possibly eliminate this problem is discussed elsewhere (Zaroubi 1996a; 1996b).

## Acknowledgments

We thank Naoshi Sugiyama and Martin White for providing the COBE normalizations of the various models. We acknowledge stimulating discussions with Marc Davis, Ravi Sheth, Douglas Scott and Idit Zehavi. This work is supported in part by US-Israel Binational Science Foundation grants 92-00355 and 94-00185, by Israel Science Foundation grants 469/92 and 590/94, and by US National Science Foundation grant PHY-91-06678.

## REFERENCES

Bertschinger E. & Dekel A., 1989, Astrophys. J., 336, L5

Crittenden R., Bond J.R., Davis R.L., Efstathiou G. & Steinhardt P.J., 1993, Phys. Rev. Lett, 61, 324

Dekel A., 1994 Ann. Rev. of Astron. & Astrophys., 32, 371

Dekel A., 1996, to appear in Proceeding of the XXXth Moriond Meeting "Clustering in the Universe".

Dekel A., Bertschinger E. & Faber S.M., 1990, Astrophys. J., 364, 349

Dekel A. & Rees M.J., 1987, Nature, 326, 455

Dressler A., 1980, Astrophys. J., 236, 351

Efstathiou G., Bond J.R. & White S.D.M., 1992, M.N.R.A.S., 258, 1p

Górski K.M., 1988, Astrophys. J., 332, L7

Górski K.M., Ratra B., Sugiyama N. & Banday A.J., 1995, Astrophys. J., 444, L65

Jaffe & Kaiser N., 1994, preprint (astro-ph/9408046)

Kaiser N., 1987, M.N.R.A.S., 227, 1

Kolatt T. & Dekel A. 1996, Astrophys. J., submitted.

Kolatt T., Dekel A., Ganon G. & Willick J. 1996, Astrophys. J., 458, 419

Peacock J.A. & Dodds S.J., 1994, M.N.R.A.S., 267, 1020

Press W.H., Teukolsky S.A., Vetterling W.T, & Flannery B.P., 1992, "Numerical Recipes" (2d ed.; Cambridge: Cambridge University Press)

Seljak, U. & Bertschinger, E. 1994, Astrophys. J., 427, 523

Sugiyama N., 1995, Astrophys. J. (Supp.), 100, 281

Turner M.S., 1993, Phys. Rev., D48, 5302

White M. & Bunn E.F., 1995, Astrophys. J., 450, 477

White M., Scott D., Silk J. & Davis M., 1995, M.N.R.A.S., 276, 69P

White S.D.M., Efstathiou G. & Frenk C.S., 1993, M.N.R.A.S., 262, 1023

Willick J.A., Courteau S., Faber S.M., Burstein D. & Dekel A., 1995, Astrophys. J., 446, 12

Willick J.A., Courteau S., Faber S.M., Burstein D., Dekel A. & Kolatt, T., 1996a, Astrophys. J., 457, 460

Willick J.A., Courteau S., Faber S.M., Burstein D. & Dekel A., 1996b, Astrophys. J. (Supp.), in preparation

Zaroubi S., 1996a, to appear in Proceeding of the XXXth Moriond Meeting "Clustering in the Universe" (astro-ph/9505103)

Zaroubi S., 1996b, in preparation

Zaroubi S. & Hoffman Y., 1996, Astrophys. J., (in press)

Zaroubi S., Hoffman Y. & Dekel, A., 1996, preprint

Zaroubi S., Hoffman Y., Fisher K.B. & Lahav O., 1995, Astrophys. J., 449, 446